\documentclass[aps,prl,twocolumn]{revtex4}

\usepackage[english]{babel}
\usepackage{graphicx}
\usepackage{amssymb}
\usepackage{amsmath, amsthm, amssymb}
\usepackage{amsthm}
\usepackage{colordvi}

\begin{document}

\title{Equation of state of a weakly interacting two-dimensional Bose gas studied at zero temperature by means of quantum Monte Carlo methods}

\author{
G.E.~Astrakharchik$^a$, J.~Boronat$^a$, J.~Casulleras$^a$, I.L.~Kurbakov$^b$, and Yu.E.~Lozovik$^b$}
\affiliation{
$^a$ Departament de F\'{\i}sica i Enginyeria Nuclear, Campus Nord B4-B5, Universitat Polit\`ecnica de Catalunya, E-08034 Barcelona, Spain\\
$^b$ Institute of Spectroscopy, 142190 Troitsk, Moscow region, Russia}

\date{\today}

\begin{abstract}
The equation of state of a weakly interacting two-dimensional Bose gas is studied at zero temperature by means of quantum Monte Carlo methods. Going down to as low densities as $na^2 \propto 10^{-100}$ permits us for the first time to obtain agreement on beyond mean-field level between predictions of perturbative methods and direct many-body numerical simulation, thus providing an answer to the fundamental question of the equation of state of a two-dimensional dilute Bose gas in the universal regime ({\it i.e.} entirely described by the gas parameter $na^2$). We also show that the measure of the frequency of a breathing collective oscillation in a trap at very low densities can be used to test the universal equation of state of a two-dimensional Bose gas.
\end{abstract}


\maketitle

The calculation of properties of weakly-interacting quantum gases was historically a very important and productive task as it led to the development of perturbative methods, such as the Bogoliubov diagonalization approach, Feynman diagrams, {\it etc}.  % Beliaev theory?
The equation of state of a three-dimensional Bose gas dates back to 1957 when the beyond mean-field correction was obtained by Lee, Huang, and Yang \cite{Lee57b}.
In the beginning of 1960s the equation of state of a dilute one-dimensional Bose gas was obtained by Girardeau \cite{Girardeau60}. From this perspective it is rather surprising that even until now the equation of state of a dilute two-dimensional (2D) gas remains an open question.
%The peculiarity of two-dimensional world is that so far there is no clear understanding of the equation of state of a weakly-interacting Bose gas.
In the last thirty years, the problem has been addressed in different studies\cite{Schick71, Popov72,Lozovik78,Hines78,Fisher88, Kolomeisky92, Ovchinnikov93,Stoof93,Cherny01,Lieb01,Andersen02,Mora03,Pricoupenko04}
%D. R. Nelson, Phys. Rev. Lett. 60, 1973 (1988)
%D. R. Nelson and H. S. Seung, Phys. Rev. B 39, 9153 (1989)
%D. R. Nelson, J. Stat. Phys. 57, 511 (1989)
but the results obtained are often incomplete and even contradictory, which indicates that the two-dimensional problem is extremely complicated. Indeed, a peculiarity of two-dimensional systems is that the coupling constant depends on the density, contrarily to three- and one- dimensional cases, where it is entirely defined by the $s$-wave scattering length $a$ and the particle mass $m$. This feature makes the analytical derivation of the equation of state much more involved.

When perturbative theories fail one can resort to {\it ab initio} numerical simulations. The correctness of perturbative equations of ground state has been checked in three dimensions \cite{Giorgini99} and in one dimension (see, {\it e.g.} %\cite{Astrakharchik04d}) %Super TG
Ref.~\cite{Astrakharchik03}) % LL
using diffusion Monte Carlo (DMC) methods. Analogous attempts\cite{Pilati05,Astrakharchik07} to find numerical agreement with beyond mean-field terms in 2D systems have not succeeded as in this case the expansion parameter has logarithmic dependence on the gas parameter $na^2$. In particular, densities as low as $na^2\propto 10^{-5}$ \cite{Mazzanti05}, $10^{-7}$ \cite{Pilati05} and $10^{-6}$ \cite{Astrakharchik07} have been reached. Our estimation below shows that for confident testing of the beyond mean-field (BMF) terms the density should be smaller than $10^{-69}$, which obviously is an extremely challenging task. In this work we do an effort towards calculations at considerably smaller densities in order to test different expressions for beyond mean-field terms present in the literature.

We use the diffusion Monte Carlo (DMC) method to address the problem. DMC solves stochastically the Schr\"odinger equation providing exact results for the ground state energy within controllable statistical errors. The construction of the guiding wave function is as in Ref.~\cite{Astrakharchik07b}. In Table~\ref{table:E} we report the energy per particle in the thermodynamic limit as a function of the density. The scattering at low densities is {\it universal}: it is independent of a particular choice of interaction potential, and is described by a single parameter, the $s$-wave scattering length. Although the calculations are done for the dipolar interaction there is no difference between results for different interaction potentials for densities $nr_0^2\ll 10^{-10}$. For convenience we keep the name of ``dipoles'' in order to distinguish between different series of calculations.

\begin{table}
\begin{tabular}{||l|ll|l|l||}
\hline
$nr_0^2$&$E/N$&$\quad$&$nr_0^2$&$E/N$\\
\hline
$2^{-4} $&$0.23338(9)              $&&$2^{-24}$&$2.6070(10){\cdot}10^{-8}$\\
$2^{-5} $&$0.095917(38)            $&&$2^{-25}$&$1.2402(5){\cdot}10^{-8}$\\
$2^{-6} $&$0.039924(16)            $&&$2^{-26}$&$5.9130(24){\cdot}10^{-9}$\\
$2^{-7} $&$0.016817(7)             $&&$2^{-27}$&$2.8268(11){\cdot}10^{-9}$\\
$2^{-8} $&$0.0071709(28)           $&&$2^{-28}$&$1.3536(5){\cdot}10^{-9}$\\
$2^{-9} $&$0.0030943(12)           $&&$2^{-29}$&$6.4953(26){\cdot}10^{-10}$\\
$2^{-10}$&$0.0013491(5)            $&&$2^{-30}$&$3.1218(12){\cdot}10^{-10}$\\
$2^{-11}$&$5.9355(24){\cdot}10^{-4}$&&$2^{-31}$&$1.5029(6){\cdot}10^{-10}$\\
$2^{-12}$&$2.6408(10){\cdot}10^{-4}$&&$2^{-32}$&$7.2448(29){\cdot}10^{-11}$\\
$2^{-13}$&$1.1846(5){\cdot}10^{-4} $&&$3{\cdot}10^{-11}$&$8.4396(51){\cdot}10^{-12}$\\
$2^{-14}$&$5.3611(21){\cdot}10^{-5}$&&$3{\cdot}10^{-12}$&$7.6254(46){\cdot}10^{-13}$\\
$2^{-15}$&$2.4421(10){\cdot}10^{-5}$&&$10^{-13}$&$2.2249(13){\cdot}10^{-14}$\\
$2^{-16}$&$1.1206(4){\cdot}10^{-5} $&&$10^{-15}$&$1.9048(11){\cdot}10^{-16}$\\
$2^{-17}$&$5.1710(21){\cdot}10^{-6}$&&$10^{-17}$&$1.6659(10){\cdot}10^{-18}$\\
$2^{-18}$&$2.3987(10){\cdot}10^{-6}$&&$10^{-20}$&$1.4028(3){\cdot}10^{-21}$\\
$2^{-19}$&$1.1179(4){\cdot}10^{-6} $&&$10^{-25}$&$1.1116(2){\cdot}10^{-26}$\\
$2^{-20}$&$5.2340(21){\cdot}10^{-7}$&&$10^{-33}$&$8.3539(34){\cdot}10^{-35}$\\
$2^{-21}$&$2.4604(10){\cdot}10^{-7}$&&$10^{-50}$&$5.4752(22){\cdot}10^{-52}$\\
$2^{-22}$&$1.1604(5){\cdot}10^{-7} $&&$10^{-67}$&$4.0746(16){\cdot}10^{-69}$\\
$2^{-23}$&$5.4925(22){\cdot}10^{-8}$&&$10^{-100}$&$2.7251(10){\cdot}10^{-102}$\\
\hline
\end{tabular}
\caption{DMC energy per particle for dipolar interaction as a function of the density $nr_0^2$ obtained by extrapolating the results for $N$ = 100 --- 800 particles to the thermodynamic limit. Conversion to units of $s$-wave scattering length $a$ can be easily done according to the relation $a=3.17222 r_0$. The energy is given in units of $\hbar^2/mr_0^2$. Statistical errors on the last digit are shown in parentheses.}
\label{table:E}
\end{table}

The leading contribution in the 2D equation of state is described by mean-field Gross-Pitaevskii (GP) theory in which all particles are considered to be in the condensate. The chemical potential is proportional to the density $\mu = g_{\text{2D}} n$ and the coefficient of proportionality is the coupling constant. %: $E^{\text{MF}}/N = g_{2D} n/2$.
Unlike three- and one- dimensional systems, where the coupling constant is independent of the density, in a two-dimensional system $g_{\text{2D}}$ itself depends on density making the construction of a precise theory very complicated. The leading contribution to the purely 2D coupling constant was first obtained by Schick \cite{Schick71} in 1971, who made use of the Beliaev method\cite{Beliaev58}. Contrary to three- and one- dimensional cases, the dependence of the coupling constant on the $s$-wave scattering length $a$ is very weak. Indeed, $a$ enters under the logarithm,
\begin{eqnarray}
g^{\text{MF}}_{\text{2D}} = \frac{4\pi\hbar^2}{m}\frac{1}{|\ln na^2|}.
\label{Emf}
\end{eqnarray}
Recently, it has been shown that a rigorous derivation of a two-dimensional Gross-Pitaevskii theory leads to the same result\cite{Lieb01}.

Before discussing beyond mean-field terms it is necessary to note that the usual mean-field relation between energy and chemical potential is accurate only in the leading term, while in three- and one- dimensional systems the MF relation is exactly linear. Indeed, the chemical potential is proportional to the coupling constant $g_{\text{2D}}$: $\mu^{\text{MF}}_{\text{2D}} = g_{\text{2D}}n = 4\pi\hbar^2n/ |m\ln na^2|$. The energy, which is found by integrating the chemical potential, is
$E^{\text{MF}}/N = 4\pi\Gamma(0,2|\ln n a^2|)\hbar^2/(m na^4)$ where $\Gamma(a,x)$ %$= -\int_{a}^\infty t^{a-1} e^{-t} dt$
is the incomplete gamma function. The energy expansion in the dilute regime $na^2\to 0$ can be obtained from the large argument expansion of $\Gamma(0,x)$ as follows:
\begin{eqnarray}
\frac{E^{\text{MF}}}{N}
= \frac{2\pi n\hbar^2/m}{|\ln na^2| + 1/2 - 1/(4|\ln na^2|) + ...}
\label{EMFexp}
\end{eqnarray}
We see that there are differences between Eq.~(\ref{EMFexp}) and the usual MF expression $E^{\text{MF}}/N = gn/2$. The differences contain logarithmically small terms that exceed the accuracy of the mean-field theory. However, in the study of beyond MF effects terms of this order are significant.

The dominant beyond mean-field terms were obtained by Popov\cite{Popov72} in 1972 (see also his book \cite{Popov83}). He obtained a recursive expression relating chemical potential $\mu$ and density $n$. At zero temperature his expression reduces to
\begin{eqnarray}
n = \frac{m\mu}{4\pi\hbar^2}\left(\ln\frac{\varepsilon_0}{\mu}-1\right)
\label{Popov}
\end{eqnarray}
where $\varepsilon_0$ is of the order of $\hbar^2/mr_0^2$, $r_0$ being the range of the interaction potential. We write the last relation introducing an unknown coefficient of proportionality $C_1$:
$\varepsilon_0 = C_1 \hbar^2/ma^2$. By solving Eq.~(\ref{Popov}) iteratively one obtains the following expression for the chemical potential
\begin{eqnarray}
\mu^{popov} = \frac{4\pi n\hbar^2/m}{|\ln na^2|+\ln |\ln {na^2}|-\ln 4\pi+\ln C_1-1...}
\label{mupopov}
\end{eqnarray}

Mathematically, the leading beyond MF term in $\mu$ and $E/N$ is a double logarithm $\ln|\ln na^2|$, and if the density is low enough inclusion of this term should be sufficient to reproduce the energy correctly. From a practical point of view this term works at such extremely low densities; that inclusion of the double logarithm term alone might even lead to larger deviations in the energy. Indeed, the second subleading term is some constant of the order of $\ln 4\pi$. To estimate a typical density of applicability of a double-logarithm term we ask that the term $\ln|\ln na^2|$ be $m$ times larger than typical constant term $\ln 4\pi$. This leads to an estimation of a characteristic density $na^2 \approx e^{-(4\pi)^m}$. Asking for a factor $m=2$ ({\it i.e.} $50\%$ accuracy) we get an extremely low density $na^2\approx 10^{-69}$, asking for $m=3$ ({\it i.e.} $33\%$ accuracy) the density drops to a quite unrealistic number $na^2\approx 10^{-862}$. We note that the universal regime where the energy depends only on the gas parameter $na^2$ extends to much larger densities, such as $na^2\approx 10^{-6}$. This situation should be contrasted with three- or one-dimensional cases, where the universal regime is perfectly described by the first beyond MF correction. We note that problems due to a logarithmic correction arise also in four dimensions \cite{Yang08}. Another important peculiarity of two dimensional world is that second and third beyond MF corrections have different signs.
%Indeed, this can be anticipated from Eq.~(\ref{mupopov}).
Furthermore, it follows that such corrections can compensate each other at some relatively ``large'' density. %Setting $m=1$ one gets $na^2 \approx 3 \times 10^{-6}$.
In fact, this effect has been observed in a full many-body calculation\cite{Astrakharchik07} at density $na^2 \approx 1 \times 10^{-6}$. Notwithstanding the energy in this point is reproduced correctly by the mean-field theory, other properties (for example, the condensate fraction) are not described precisely.

\begin{table}
\begin{tabular}{|c|c|l|c|l|}
\hline
year&Ref.&first author&type&terms\\
\hline
1971&\cite{Schick71}&Schick&MF&$E^{\text{MF}}/N = 2\pi\hbar^2n/(m|\ln na^2|)$\\
1971&\cite{Popov72}&Popov\cite{FootNote1}&BMF&$\ln|\ln na^2| - \ln 4\pi -1/2$\\
1978&\cite{Lozovik78}&Lozovik &BMF&$\ln|\ln na^2| - \ln 4\pi +1/2$\\
1978&\cite{Hines78}&Hines&BFM&
$\ln|\ln(na^2\!/\!\pi)|\!-\!\ln 2\pi^3\!-\!2\gamma\!+\!3/2$\\
1988&\cite{Fisher88}&Fisher& BFM&$\ln|\ln na^2| - \ln 4\pi - 1/2$\\
1992&\cite{Kolomeisky92}&Kolomeisky&BFM&$\ln|\ln(4\pi na^2)| - \ln 4\pi$\\
1993&\cite{Ovchinnikov93}&Ovchinnikov&BMF&$\ln|\ln na^2|$\\
2001&\cite{Lieb01}&Lieb&MF&$E^{\text{MF}}/N = 2\pi\hbar^2n/(m|\ln na^2|)$\\
2001&\cite{Cherny01}&Cherny&BMF&$\ln|\ln na^2| - \ln\pi -2\gamma -1/2$\\
2002&\cite{Andersen02}&Andersen&BMF&$\ln|\ln na^2| - \ln 4\pi -1/2$\\
2003&\cite{Mora03}&Mora&BMF&$\ln|\ln na^2| - \ln\pi -2\gamma-1/2$\\
2004&\cite{Pricoupenko04}&Pricoupenko&BMF&$\ln|\ln na^2| - \ln\pi -2\gamma-1/2$\\
2005&\cite{Pilati05}&Pilati&BMF&$0.86\ln |\ln na^{2}|-2.26$\\
%1993&&H.T.C. Stoof and M. Bijlsma, Phys. Rev. E 47, 939\\ - cite for discussion of applicability of Popov's theory
%D. R. Nelson, Phys. Rev. Lett. 60, 1973 (1988)
%D. R. Nelson and H. S. Seung, Phys. Rev. B 39, 9153 (1989)
%D. R. Nelson, J. Stat. Phys. 57, 511 (1989)
\hline
\end{tabular}
\caption{Literature overview: equation of state of a dilute two-dimensional Bose gas.
Fifth column, terms of the low-density expansion appearing as straight lines in Fig.~\ref{Fig1} in the limit of large $\ln|\ln na^2|$.}
\label{table:theory}
\end{table}

The complexity of the problem explains the large number of works dedicated to the analytical study of the equation of state of dilute 2D Bose gas. A summary of these works is presented in Table~\ref{table:theory}, where we report the expression for the energy per particle. Almost all beyond MF theories agree on the presence of the $\ln|\ln na^2|$ term but differ in the constant, or explicitly discuss that the constant term can not be obtained within the used approach. Previous diffusion Monte Carlo calculations \cite{Pilati05,Astrakharchik07} have shown the existence of a universal equation of state, but even at the smallest considered densities none of the equations of state given in Table~\ref{table:theory} was able to reproduce the energy correctly.

\begin{figure}
\begin{center}
\includegraphics[angle=-90,width=\columnwidth]{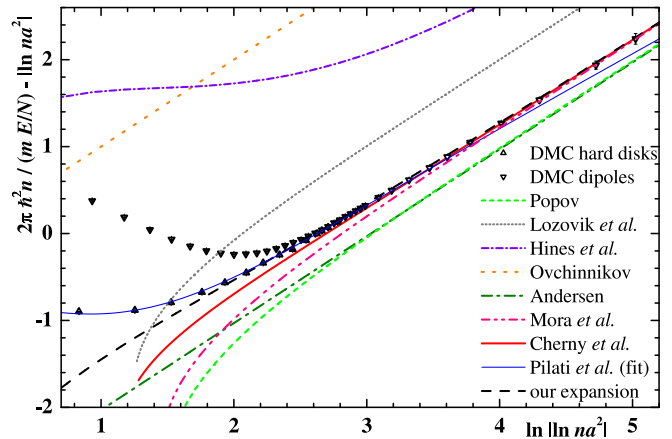}
\caption{(Color online) Non-universal beyond MF corrections $2\pi\hbar^2n/[mE/N]-|\ln na^2|$ in the energy per particle as a function of the double logarithm of the gas parameter $\ln|\ln na^2|$. Symbols, DMC results: up triangles, hard disks\cite{Pilati05}, down triangles dipoles. Lines, different equations of state, see Table~\ref{table:theory}. Black dashed line, Eq.~(\ref{Ean}).}
\label{Fig1}
\end{center}
\end{figure}

In the present DMC calculation at ultra-low densities we reach the regime where the analytical expansions from Table~\ref{table:theory} are applicable and therefore it is possible to test them. According to the majority of theories the complete expression for the energy can be generalized as $E/N = 2\pi\hbar^2n/[m(|\ln na^2|+\ln|\ln na^2| - const+...)]$. Then, the beyond MF terms should be tested by inverting the energy and subtracting the MF term, according to the combination $2\pi\hbar^2n/[mE/N]-|\ln na^2| = \ln|\ln na^2| - const$. Figure~\ref{Fig1} shows DMC results of this combination as a function of the double logarithm of the gas parameter $\ln|\ln na^2|$ for hard disk and dipolar interaction potentials. The analytical results from Table~\ref{table:theory} are drawn with lines. The majority of approaches predict the first beyond mean-field term as $\ln|\ln na^2|$ that would correspond to a linear behavior with unitary slope in Fig.~\ref{Fig1}. The only two exceptions are the theory of Hines {\it et al.} \cite{Hines78} and a fit to DMC results of hard disks \cite{Pilati05}. The region of applicability of the latter fit is limited to $1.5<\ln|\ln na^2|<2.8$. Our results for dipolar interaction potential are compatible with the linear dependence and unitary slope (for the largest $x=\ln|\ln na^2|\gtrsim 4$ the numerical simulations are very difficult and the error $\Delta E$ in the estimation of the energy is exponentially amplified in the quantity of interest as $(n \Delta E)/E^2\approx e^x (\Delta E)/E$). The third beyond mean-field correction corresponds to a constant shift in Fig.~\ref{Fig1}. Here there are the major discrepancies between theories. The reason for that is the complexity of perturbation theory in two-dimensions. One has to evaluate correctly all the terms contributing to a given level of accuracy. Just to name some sources for this constant, already from the recursive expression (\ref{Popov}) one sees that the self-consistent evaluation of $\mu$ contributes to its value. Also the effect of the quantum fluctuations (depletion of the condensate density), the difference between cut-off length, the range of the potential and the $s$-wave scattering length contribute on this level. The Euler's gamma constant $\gamma$ appearing in some of the theoretical results is related to the dependence in the ladder approximation of the chemical potential on the scattering amplitude\cite{Beliaev58}, which in turn contains $\gamma$ in its short-range expansion\cite{Hines78}.

Our DMC results are in agreement within error bars with the constant shift $-\ln\pi -1/2-2\gamma$ obtained in Refs.~\cite{Cherny01,Mora03,Pricoupenko04} for $\ln|\ln na^2|\gtrsim 3$. Cherny and Shanenko do expansion in terms of a dimensionless in-medium scattering amplitude $u$ obeying the equation $1/u+\ln u=-\ln(\pi na^2)-2\gamma$. The dimensionless energy $\varepsilon=Em/(2\pi\hbar^2nN)$ is then expanded as a series $\varepsilon = u +u^2/2-u^3+...$ We test the accuracy of this description and find that the agreement with the first two terms of this expansion is notably good. The deviations enter on the level of the third beyond MF term. We apply a fitting procedure to find the coefficient in front of $u^3$ with the $\chi^2$ criterion; the result is $mE/(2\pi\hbar^2Nn) = u +u^2/2 - 2.0(1) u^3$. This fit describes correctly the energy in the universal regime (where results for dipoles and the hard disks coincide) and at densities $\ln|\ln na^2|\gtrsim 2$ for hard disks. We note that the hard disk interaction potential is completely described by only one parameter and its equation of state is expected to be the most universal.

In terms of the gas parameter our best perturbative description for the chemical potential is given by
\begin{eqnarray}
\mu= \frac{4\pi\hbar^2n/m}{
     |\ln na^2|
+ \ln|\ln na^2| + C^{\mu}_1
+\frac{\ln|\ln na^2| +C^{\mu}_2}{|\ln na^2|}+...
},
\label{muan}
\end{eqnarray}
where
$C^{\mu}_1 = -\ln\pi -2\gamma -1 = - 3.30...$ and
$C^{\mu}_2 = -\ln\pi -2\gamma +2.0(1) = -0.3(1)$.
The corresponding expression for the ``universal'' energy per particle is
\begin{eqnarray}
\frac{E}{N} = \frac{2\pi\hbar^2n/m}{
     |\ln na^2|
+ \ln|\ln na^2| + C^{E}_1
+\frac{\ln|\ln na^2| + C^{E}_2}{|\ln na^2|}+...
},
\label{Ean}
\end{eqnarray}
where
$C^{E}_1 = C^{\mu}_1+1/2$ and % =  -\ln\pi -2\gamma -1/2 = - 2.80...$ and
$C^{E}_2 = C^{\mu}_2+1/4$    % =  -\ln\pi -2\gamma +2.0(1) +1/4= -0.05(10)$.
(in the mean-field case, described by Eqs.~(\ref{Emf}) and (\ref{EMFexp}), relation between $C^{E}_1$ and $C^{\mu}_1$ is similar, while relation between $C^{E}_2$ and $C^{\mu}_2$ is different due to double-logarithm term present in Eq.~(\ref{Ean})).

\begin{figure}
\begin{center}
\includegraphics[angle=-90,width=\columnwidth]{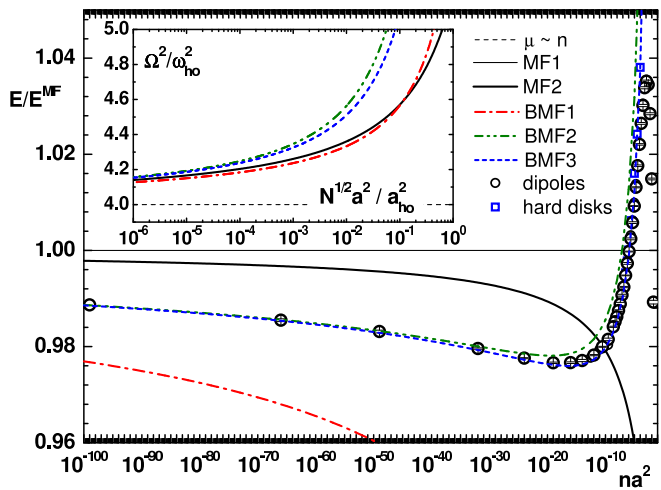}
\caption{(Color online) Main plot: energy in units of MF energy $E^{\text{MF}}_1/N = 2\pi\hbar^2 n/|m\ln na^2|$ as a function of the gas parameter, inset : square of the lowest breathing mode frequency, $\Omega^2$ as a function of the LDA parameter $N^{1/2}r_0^2/a_{\text{ho}}^2$.
Thin dashed line (inset), mean-field linear equation of state $\mu^{\text{MF}}_{0} \propto n$,
thin solid line (inset and main plot): mean-field expression of Schick for the energy $E^{\text{MF}}_1$,
thick solid line (main plot): mean-field expression
%$E^{\text{MF}}_2 = - 4\pi\hbar^2 \Ei(2\ln n a^2)\;/\;m$,
$E^{\text{MF}}_2/N = 4\pi\Gamma(0,2|\ln n a^2|)\hbar^2/(m na^4)$
dash line with one dot: first beyond MF correction,
dash line with two dots, second BMF correction,
short-dashed line: three BFM terms,
circles: DMC results for dipoles,
squares: DMC results for hard disks.
}
\label{Fig2}
\end{center}
\end{figure}

We illustrate the convergence of the series in Fig.~\ref{Fig2} where we plot the quotient between the energy and mean-field prediction of Schick\cite{Schick71}. One sees that contrary to three- and one- dimensional cases, the mean-field prediction is not precise even at extremely dilute values of the gas parameter\cite{Astrakharchik07}. Moreover, differences between the expression of Schick and the integrated Schick's chemical potential matter at the considered densities. Inclusion of the leading beyond mean-field correction ({i.e.} double-logarithm term $\ln|\ln na^2|$) does not improve the description for densities $na^2 \gtrsim 10^{-100}$. This term is negative and it leads to an underestimated energy. The second BMF correction ($C^E_1$ and $C^\mu_1$ terms in Eqs.~(\ref{Ean},\ref{muan})) has a different sign and effectively cancels the first BMF term at the density $na^2\approx 10^{-6}$. Inclusion of this term permits to recover the DMC results up to $na^2\approx 10^{-30}$. Inclusion of the third BMF term, Eq.~(\ref{Ean}), extends the region of applicability of the series expansion up to relatively ``high'' densities $na^2 \approx 10^{-3}$.

The correctness of the series expansion of the 2D equation of state can be verified in high precision experiments measuring the frequency $\Omega$ of the lowest breathing mode. We note that the presence of a tight transverse harmonic confinement modifies the coupling constant reducing the effectively $s$-wave scattering length $a$ in (\ref{Emf}) by an exponentially small factor $\sim\exp\{-\sqrt{\pi/2}\;a_{\text{ho}}/a_{\text{3D}}\}$, where $a_{\text{ho}}$ is oscillator length and $a_{\text{3D}}$ is the $s$-wave scattering length \cite{Petrov00b}. This makes ultradilute 2D densities be experimentally accessible exploiting confinement induced resonance. We use local density approximation (LDA) to describe a trapped gas and show our predictions for $\Omega$ in the inset of Fig.~\ref{Fig2}. As one can see, the effect is large and therefore the observation of $\Omega$ provides a sensitive experimental tool for testing the equation of state.

To conclude, we study the zero temperature equation of state of weakly interacting 2D Bose gas down to ultra dilute densities $na^2 \propto 10^{-100}$ and confront the obtained energy with a large number of different analytical expressions for the energy. All theories agree at mean-field level and the majority agrees on the first beyond MF term, which, still does not provide an accurate quantitative description even at such low densities. A good agreement between perturbative series and results of the present {\it ab initio} calculations is found for the first time. The third beyond MF term is fitted and the corresponding equation of state is explicitly written in terms of the gas parameter. Finally, we suggest future experimental work to measure the frequency of a breathing collective oscillation in a trap at very low densities since our results prove that this frequency is a very sensitive tool for verifying the universal 2D equation of state.

The work was partially supported by (Spain) Grant No. FIS2005-04181, Generalitat de Catalunya Grant No. 2005SGR-00779 and RFBR. G.E.A. acknowledges post doctoral fellowship by MEC (Spain).

%\bibliography{astra}

\end{document}